\documentstyle[aps,prl,multicol,epsf]{revtex}

\begin{document}

\draft

\begin{multicols}{2}
\narrowtext

{\large\bf Comment on "Macrospopic Equation for the Roughness of
Growing Interfaces in Quenched Disorder"}

\medskip

In a recent Letter \cite{braun} Braunstein and Buceta 
introduced a 'macroscopic' equation for the time evolution of 
the width of interfaces belonging to the
directed percolation depinning (DPD) universality class \cite{dpd}.
From numerical simulations of the DPD model, they inferred
an {\em ansatz} (Eq.(1) in Ref.\cite{braun}) 
for the time derivative of 
the interface width (called DSIW in Ref.\cite{braun}) at the
depinning transition. Braunstein and Buceta found 
that their formula fitted the numerical data at the depinning trasition, 
for $q_c=0.539$ and $\beta = 0.63$,
with the appropriate election of some arbitrary constants. 
 
Here we argue that, contrary to what it is claimed 
in Ref.\cite{braun}, Braunstein and Buceta's formula 
does not describe the 'macroscopic' behaviour 
of the interface. The formula proposed in Ref\cite{braun} for 
the DSIW is an approximation to the very short times regime (when less
than one layer has been completed), which is not significant for the 
description of the surface dynamics at large scales.
We obtain analitically the short time
behaviour of the DPD model, which is valid for any q and explains
the apperance of an exponential term in the formula of Ref.\cite{braun}
for the DSIW. 

Let us consider the DPD model in a system of size $L$ 
and a density $q$ of blocked cells ($p=1-q$ density of free cells).
We are interested in the very short times regime when the first monolayer
still has not been completed, {\it i.e.} the number of
growth attempts $N$ is $N \ll L$ (this corresponds to times $t=N/L \ll 1$). 
In this regime, the probability of having a column $i$ with height 
$h_i > min(h_{i-1},h_{i+1}) +2$ is negligible and 
the columns are growing almost independently. The growth 
at this early stage can be seen as a random deposition (RD)
process \cite{alb} in which
every column grows in one unit with probability $p/L$. The short time
regime of the DPD model is then like RD, which is solvable exactly,
but with the additional ingredient of a density $q$ of blocked sites.

One can see that, within this approximation,  
the probability of having a column with height $h$ after $N$ 
growth attempts is given by
\begin{equation}
\label{prob_h}
P(N,h) = \frac{(Nsp)^h}{h!} e^{-Ns} 
+ q p^h \sum_{r=h+1}^{N} \frac{(Ns)^r}{r!} e^{-Ns}, 
\end{equation}
where $s=1/L$ is the probability of attempting to growth a column and the
usual approximation 
$s^r (1-s)^{N-r} \: N!/[(N-r)!r!] \approx (Ns)^r exp(-Ns)/r!$
has been made.

From the probability (\ref{prob_h}), one can calculate the interface
width $W^2 = \langle h^2\rangle - \langle h \rangle^2 $
and then the time derivative, which leading terms are
\begin{equation}
\label{width}
\frac{dW^2}{dt} = p e^{-qt} + 2 p^2 e^{-qt}\left(\frac{e^{-qt}-1}{q} + t 
\right),
\end{equation} 
where $t = Ns = N/L$ is the time in the units used in Ref.\cite{braun}.
This formula gives the exact time evolution of $\frac{dW^2}{dt}$
for any $q$ (not only at $q_c=0.539$) 
and is valid for times $t \ll 1$. 
For times $t > 1$ differences between neighbouring columns are 
likely to be larger than $2$ resulting in horizontal 
correlations and the break down of (\ref{width}). 
A comparison of Eq.(\ref{width}) with
numerical simulations of the DPD model is presented in Figure 1. 

Our calculation suggests that the exponential term in the {\em ansatz}
of Ref.\cite{braun} is actually produced by the usual 
random deposition-like dynamics, which occurs in any growth
model \cite{alb} for short times.

\bigskip

\noindent Juan M. L\'opez,\\
\noindent Department of Mathematics, Imperial College,\\ 
\noindent 180 Queen's Gate, London SW7 2BZ, United Kingdom\\

\noindent Jos\'e J. Ramasco$^{*,\dag}$ and Miguel A. Rodr{\'\i}guez$^\dag$\\ 
\noindent $^*$Departamento de F\'{\i}sica Moderna,\\
\noindent Universidad de Cantabria, Avenida Los Castros s/n,\\
\noindent Santander E-39005, Spain 

\noindent $^\dag$Instituto de F\'{\i}sica de Cantabria,\\ 
\noindent Consejo Superior de Investigaciones Cient\'\i ficas -- 
Universidad de Cantabria, \\
\noindent Santander E-39005, Spain

\begin{figure}
\centerline{
\epsfxsize=6cm
\epsfbox{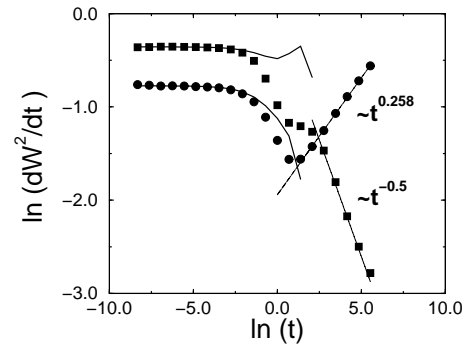}}
\caption{Numerical results for the DPD model in a
system of size $L=2^{13}$
for $q_c = 0.539$ (circles) and $q = 0.3$ (squares).
Continuous lines correspond to Eq.(2) and fit the data for $t \ll 1$.
For larger times our approximation is not valid any longer and
the power law $t^{2\beta-1}$ takes over with
$\beta = 0.623$ and $\beta=0.3$ for $q_c = 0.539$ 
and $q = 0.3$ respectively
(dotted lines).}
\end{figure}

\end{multicols}

\end{document}